# Hierarchical organization of modularity in metabolic networks

E. Ravasz[1], A.L. Somera[2], D.A. Mongru[2], Z.N. Oltvai[2] & A.-L. Barabási[1]

[1]Department of Physics, University of Notre Dame, Notre Dame, IN  46556

[2]Department of Pathology, Northwestern University, Chicago, IL  60611

**Spatially or chemically isolated functional modules composed of several cellular components and carrying discrete functions are considered fundamental building blocks of cellular organization, but their presence in highly integrated biochemical networks lacks quantitative support.  Here we show that the metabolic networks of 43 distinct organisms are organized into many small, highly connected topologic modules that combine in a hierarchical manner into larger, less cohesive units, their number and degree of clustering following a power law. Within *Escherichia coli* the uncovered hierarchical modularity closely overlaps with known metabolic functions. The identified network architecture may be generic to system-level cellular organization.**

The identification and characterization of system-level features of biological organization is a key issue of post-genomic biology (*1, 2, 3*). The concept of modularity assumes that cellular functionality can be seamlessly partitioned into a collection of modules. Each module is a discrete entity of several elementary components and performs an identifiable task, separable from the functions of other modules (*1, 4, 5, 6, 7, 8*). Spatially and chemically isolated molecular machines or protein complexes (such as ribosomes and flagella) are prominent examples of such functional units, but more extended modules, such as those achieving their isolation through the initial binding of a signaling molecule (*9*) are also apparent.

Simultaneously, it is now widely recognized that the thousands of components of a living cell are dynamically interconnected, so that the cell's functional properties are ultimately encoded into a complex intracellular web of molecular interactions (*2, 3, 4, 5, 6, 8*). This is perhaps most evident when inspecting cellular metabolism, a fully connected biochemical network in which hundreds of metabolic substrates are densely integrated via biochemical reactions. Within this network, however, modular organization (i.e., clear boundaries between sub-networks) is not immediately apparent. Indeed, recent studies have demonstrated that the probability that a substrate can react with $k$ other substrates (the degree distribution $P(k)$ of a metabolic network) decays as a power law $P(k) \sim k^{-\nu}$ with $\nu \approx 2.2$ in all organisms (*10, 11*) suggesting that metabolic networks have a scale-free topology (*12*). A distinguishing feature of such scale-free networks is the existence of a few highly connected nodes (e.g., pyruvate or CoA), which participate in a very large number of metabolic reactions. With a large number of links, these hubs integrate all substrates into a single, integrated web in which the existence of fully separated modules is prohibited by definition (Fig. 1a).

Yet, the dilemma of a modular- versus a highly integrated module-free metabolic network organization remains. A number of approaches for analyzing the functional capabilities of metabolic networks clearly indicate the existence of separable functional elements (*13, 14*). Also, from a purely topologic perspective the metabolic network of *Escherichia coli* is known to possess a high clustering coefficient (*11*), a property that is suggestive of a modular organization (see below). In itself, this implies that the metabolism of *E. coli* has a modular topology, potentially comprising several densely interconnected functional modules of varying sizes that are connected by few inter-module links (Fig. 1b). However, such clear-cut modularity imposes severe restrictions on the degree distribution, implying that most nodes have approximately the same number of links, which contrasts with the metabolic network's scale-free nature (*10, 11*).

To determine if such a dichotomy is indeed a generic property of all metabolic networks we first calculated the *average clustering coefficient* for 43 different organisms (*15, 16*) as a function of the number of distinct substrates, $N$, present in their metabolism. The clustering coefficient, defined as $C_i = 2n/k_i(k_i-1)$, where $n$ denotes the number of direct links connecting the $k_i$ nearest neighbors of node $i$ (*17*), is equal to one for a node at the center of a fully inter-linked cluster, while it is zero for a metabolite that is part of a loosely connected group (Fig. 2a). Therefore, $C_i$ averaged over all nodes $i$ of a metabolic network is a measure of the network's potential modularity. We find that for all 43 organisms the clustering coefficient is about an order of magnitude larger than that expected for a scale-free network of similar size (Fig. 2b), suggesting that metabolic networks in all organisms are characterized by a high intrinsic potential modularity. We also observe that in contrast with the prediction of the scale-free model, for which the clustering coefficient decreases as $N^{-0.75}$ (*18*), the clustering coefficient of metabolic networks' is independent of their size (Fig. 2b).

Taken together, these results demonstrate a fundamental conflict between the predictions of the current models of metabolic organization. The high, size independent clustering coefficient offers strong evidence for modularity, while the power law degree distribution of all metabolic networks (*10, 11*) strongly support the scale-free model and rule out a manifestly modular topology. To resolve this apparent contradiction we propose a simple heuristic model of metabolic organization, which we refer to as a "hierarchical" network (Fig. 1c) (*19*). In such a model network, our starting point is a small cluster of four densely linked nodes. Next we generate three replicas of this hypothetical module and connect the three external nodes of the replicated clusters to the central node of the old cluster, obtaining a large 16-node module. Subsequently, we again generate three replicas of this 16-node module, and connect the peripheral nodes to the central node of the old module (Fig. 1c). These replication and connection steps can be repeated indefinitely, in each step quadrupling the number of nodes in the system. The architecture of such a network integrates a scale-free topology with an inherent modular structure. It has a power law degree distribution with degree exponent $\nu = 1 + \ln4/\ln3 = 2.26$, in agreement with the $\nu = 2.2$ observed in metabolic networks. Its clustering coefficient, $C \approx 0.6$, is also comparable with that observed for metabolic networks. Most importantly, the clustering coefficient of the model is independent of the size of the network, in agreement with the results of Fig. 2b.

A unique feature of the proposed network model, not shared by either the scale-free (Fig. 1a) or modular (Fig. 1b) models, is its hierarchical architecture. This hierarchy, which is evident from a visual inspection, is intrinsic to the assembly by repeated quadrupling of the system. The hierarchy can be characterized quantitatively by using the recent observation (*20*) that in deterministic scale-free networks the clustering coefficient of a node with $k$ links follows the scaling law $C(k) \sim k^{-1}$. This scaling law quantifies the coexistence of a hierarchy of nodes with different degrees of modularity, as measured by the clustering coefficient, and is directly relevant to our model (Fig. 1c). Indeed, the nodes at the center of the numerous 4-node modules have a clustering coefficient $C=3/4$. Those at the center of a 16-node module have $k = 13$ and $C = 2/13$, while those at the center of the 64 node modules have $k = 40$ and $C = 2/40$, indicating that the higher a node's connectivity the smaller is its clustering coefficient, asymptotically following the $1/k$ law.

To investigate if such hierarchical organization is present in cellular metabolism we measured the $C(k)$ function for the metabolic networks of all 43 organisms. As shown in Fig. 2c-f, for each organism $C(k)$ is well approximated by $C(k) \sim k^{-1}$, in contrast to the $k$-independent $C(k)$ predicted by both the scale-free and modular networks. This provides direct evidence for an inherently hierarchical organization. Such hierarchical modularity reconciles within a single framework all the observed properties of metabolic networks: their scale-free topology; high, system size independent clustering coefficient and the power-law scaling of $C(k)$.

A key issue from a biological perspective is whether the identified hierarchical architecture reflects the true functional organization of cellular metabolism. To uncover potential relationships between topological modularity and the functional classification of different metabolites we concentrate on the metabolic network of *Escherichia coli*, whose metabolic reactions have been exhaustively studied, both biochemically and genetically (*21*). With a previously established graph-theoretical representation in hand (*10*), we first subjected *E. coli*'s metabolic organization to a three step reduction process, replacing non-branching pathways with equivalent links, allowing us to decrease its complexity without altering the network topology (*16*). Next, we calculated the topological overlap matrix, $O_T(i,j)$, of the condensed metabolic network (Fig. 3a). A topological overlap of one between substrates *i* and *j* implies that they are connected to the same substrates, while a zero value indicates that *i* and *j* do not share links to common substrates among the metabolites they react with. The metabolites that are part of highly integrated modules have a high topological overlap with their neighbors, and we find that the larger the overlap between two substrates within the *E. coli* metabolic network the more likely it is that they belong to the same functional class.

As the topological overlap matrix is expected to encode the comprehensive functional relatedness of the substrates forming the metabolic network, we investigated whether potential functional modules encoded in the network topology can be uncovered automatically. Initial application of an average-linkage hierarchical clustering algorithm (*22*) to the overlap matrix of the small hypothetical network shown in Fig. 3a placed those nodes that have a high topological overlap close to each other (Fig. 3b). Also, the method has clearly identified the three distinct modules built into the model of Fig. 3a, as illustrated by the fact that the *EFG* and *HIJK* modules are closer to each other in a topological sense than the *ABC* module (Fig. 3b). Application of the same technique on the *E. coli* overlap matrix $O_T(i,j)$ provides a global topologic representation of *E. coli* metabolism (Fig. 4a). Groups of metabolites forming tightly interconnected clusters are visually apparent, and upon closer inspection the hierarchy of nested topologic modules of increasing sizes and decreasing interconnectedness are also evident. To visualize the relationship between topological modules and the known functional properties of the metabolites, we color coded the branches of the derived hierarchical tree according to the predominant biochemical class of the substrates it produces, using the standard, small molecule biochemistry based classification of metabolism (*15*). As shown in Fig. 4a, and in the three dimensional representation in Fig. 4b, we find that most substrates of a given small molecule class are distributed on the same branch of the tree (Fig. 4a) and correspond to relatively well-delimited regions of the metabolic network (Fig. 4b). Therefore, there are strong correlations between shared biochemical classification of metabolites and the global topological organization of *E. coli* metabolism (Fig. 4a, bottom, and Ref. 16).

To correlate the putative modules obtained from our graph theory-based analysis to actual biochemical pathways, we concentrated on the pathways involving the pyrimidine metabolites. Our method divided these pathways into four putative modules (Fig. 4c), which represent a topologically well-limited area of *E. coli* metabolism (Fig. 4b, circle). As shown in Fig. 4d, all highly connected metabolites (Fig. 4d, red-boxes) correspond to their respective biochemical reactions within pyrimidine metabolism, together with those substrates that were removed during the original network reduction procedure, and then re-added (Fig. 4d, green boxes). However, it is also apparent that putative module boundaries do not always overlap with intuitive 'biochemistry-based' boundaries. For instance, while the synthesis of UMP from L-glutamine is expected to fall within a single module based on a linear set of biochemical reactions, the synthesis of UDP from UMP leaps putative module boundaries. Thus, further experimental and theoretical analyses will be needed to understand the relationship between the decomposition of *E. coli* metabolism offered by our topology-based approach, and the biologically relevant sub-networks.

The organization of metabolic networks is likely to combine a capacity for rapid flux reorganization with a dynamic integration with all other cellular function (*11*). Here we show that the system-level structure of cellular metabolism is best approximated by a hierarchical network organization with seamlessly embedded modularity. In contrast to current, intuitive views of modularity (Fig. 1b) which assume the existence of a set of modules with a non-uniform size potentially separated from other modules, we find that the metabolic network has an inherent self-similar property: there are many highly integrated small modules, which group into a few larger modules, which in turn can be integrated into even larger modules. This is supported by visual inspection of the derived hierarchical tree (Fig. 4a), which offers a natural breakdown of metabolism into several large modules, which are further partitioned into smaller, but more integrated sub-modules.

The mathematical framework proposed here to uncover the presence or absence of such hierarchical modularity, and to delineate the modules based on the network topology could apply to other cellular- and complex networks as well. As scale-free topology has been found at many different organizational levels, ranging from protein interaction, to genetic (*23*) and protein domain (*24*) networks, it is possible that biological networks are always accompanied by a hierarchical modularity. Some non-biological networks, ranging from the World Wide Web to the Internet, often combine a scale-free topology with a community structure (i.e., modularity) (*25, 26, 27*), therefore these networks are also potential candidates for hierarchical modularity. For biological systems hierarchical modularity is consistent with the notion that evolution may act at many organizational levels simultaneously: the accumulation of many local changes, that affect the small, highly integrated modules, could slowly impact the properties of the larger, less integrated modules. The emergence of the hierarchical topology via copying and reusing existing modules (*1*) and motifs (*8*), a process reminiscent of the results of gene duplication (*28, 29*), offers a special role to the modules that appeared first in the network. While the model of Fig. 1c reproduces the large-scale features of the metabolism, understanding the evolutionary mechanism that explains the simultaneous emergence of the observed hierarchical and scale-free topology of the metabolism, and its generality to cellular organization, is now a prime challenge.

**References and Notes**

1. L. H. Hartwell, J. J. Hopfield, S. Leibler, A. W. Murray, *Nature* **402**, C47 (1999).

2. H. Kitano, *Science* **295**, 1662-1664 (2002).

3. Y. I. Wolf, G. Karev, E. V. Koonin, *Bioessays* **24**, 105 (2002).

4. D. A. Lauffenburger, *Proc Natl Acad Sci U S A* **97**, 5031 (2000).

5. C. V. Rao, A. P. Arkin, *Annu Rev Biomed Eng* **3**, 391 (2001).

6. N. S. Holter, A. Maritan, M. Cieplak, N. V. Fedoroff, J. R. Banavar, *Proc Natl Acad Sci Sci U S A* **98,** 1693 (2001).

7. J. Hasty, D. McMillen, F. Isaacs, J. J. Collins, *Nat Rev Genet* **2**, 268 (2001).

8. S. S. Shen-Orr, R. Milo, S. Mangan, U. Alon, *Nature Genetics* **31**, 64 (2002).

9. U. Alon, M. G. Surette, N. Barkai, S. Leibler, *Nature* **397**, 168 (1999).

10. H. Jeong, B. Tombor, R. Albert, Z. N. Oltvai, A.-L. Barabási, *Nature* **407**, 651 (2000).

11. A. Wagner, D. A. Fell, *Proc R Soc Lond B Biol Sci* **268**, 1803 (2001).

12. A.-L. Barabási, R. Albert, *Science* **286**, 509 (1999).

13. C. H. Schilling, D. Letscher, B. O. Palsson, *J Theor Biol* **203**, 229 (2000).

14. S. Schuster, D. A. Fell, T. Dandekar, *Nat Biotechnol* **18**, 326 (2000).

15. R. Overbeek *et al.*, *Nucleic Acids Res* **28**, 123 (2000).

16. Supporting Online Material, available at http://www.nd.edu/~networks/cell/index.html.

17. D. J. Watts, S. H. Strogatz, *Nature* **393**, 440 (1998).

18. R. Albert, A. L. Barabási, *Rev Mod Phys* **74**, 47 (2002).

19. A.-L. Barabási, E. Ravasz, T. Vicsek, *Physica A* **299**, 559 (2001).

20. S. N. Dorogovtsev, A. V. Goltsev, J. F. F. Mendes, in press (available at http://xxx.lanl.gov/abs/cond-mat/0112143) (2001).

21. P. D. Karp, M. Riley, S. M. Paley, A. Pellegrini-Toole, M. Krummenacker, *Nucleic Acids Res* **30**, 56 (2002).

22. M. B. Eisen, P. T. Spellman, P. O. Brown, D. Botstein, *Proc Natl Acad Sci U S A* **95**, 14863 (1998).

23. D. E. Featherstone, K. Broadie, *Bioessays* **24** (3), 267 (2002).

24. S. Wuchty, *Mol Biol Evol* **18**, 1694 (2001).

25. G. W. Flake, S. Lawrence, C. L. Giles, F. M. Coetzee, *Computer* **53**, 66 (2002).

26. M. Girvan, M. E. J. Newman, *Proc. Natl. Acad. Sci. USA* **99**, 7821 (2002).

27. A. Vazquez, R. Pastor-Satorras, A. Vespignani, in press (available at http://xxx.lanl.gov/abs/cond-mat/0112400) (2001).

28. A. Vazquez, A. Flammini, A. Maritan, A. Vespignani, in press (available at http://xxx.lanl.gov/abs/cond-mat/0108043) (2001).

29. R. P. Solé, D. E. Smith, R. Pastor-Satorras, T. Kepler, Santa Fe Preprint (available at http://www.santafe.edu/sfi/publications/wpabstract/200108041).

30. **Acknowledgement.** We thank T. Vicsek and I.J. Farkas for discussions on the hierarchical model. We also thank C. Waltenbaugh and J. W. Campbell for comments on the manuscript and the WIT project for making their database publicly available. Research at the University of Notre Dame and at Northwestern University was supported by grants from the Department of Energy and the National Institute of Health Correspondence and request for materials should be addressed to A.-L.B. (e-mail: alb@nd.edu) or Z.N.O. (e-mail: zno008@northwestern.edu).

**Supporting Online Material**

www.nd.edu/~networks/cell

1. Network Models

2. Graph Theoretic Characterization of the Metabolism

3. Clustering and Functional Characterization

4. Figs. S1 to S15

**Figure Legends**

*Figure 1.* **Complex network models.**

(a) Left: A schematic illustration of a scale-free network, whose degree distribution follows a power law. In such a network a few highly connected nodes, or hubs (blue circles) play an important role in keeping the whole network together. On the right, a typical configuration of a scale-free network with 256 nodes is shown, obtained using the scale-free model, which requires adding a new node at each time such, that existing nodes with higher degrees of connectivity have a higher chance of being linked to the new nodes (*12*). The nodes are arranged in space using a standard clustering algorithm (http://vlado.fmf.uni-lj.si/pub/networks/pajek/) to illustrate the absence of an underlying modularity. (b) Left: Schematic illustration of a manifestly modular network made of four highly interlinked modules connected to each other by a few links. This intuitive topology does not have a scale-free degree distribution, as most of its nodes have a similar number of links, and hubs are absent. A standard clustering algorithm easily uncovers the network's inherent modularity (right panel) by partitioning a modular network of $N = 256$ nodes into the four isolated structures built into the system. (c) Left: The hierarchical network has a scale-free topology with embedded modularity, the increasing hierarchical levels labeled with blue, green and red, respectively. As shown on the right, standard clustering algorithms are less successful in uncovering the network's underlying modularity. For a detailed quantitative characterization of the three network models see Ref. 16.

*Figure 2.* **Evidence of hierarchical modularity in metabolic networks.**

(a) The clustering coefficient offers a measure of the degree of interconnectivity in the neighborhood of a node (*17*). For example, a node whose neighbors are all connected to each other has C = 1 (left), while a node with no links between its neighbors has C = 0 (right). (b) The average clustering coefficient, $C(N)$, for 43 organisms (*10*) is shown as a function of the number of substrates $N$ present in each of them. Species belonging to Archae (purple), Bacteria (green), and Eukaryotes (blue) are shown. The dashed line indicates the dependence of the clustering coefficient on the network size for a module-free scale-free network, while the diamonds denote *C* for a scale-free network with the same parameters (N and number of links) as observed in the 43 organisms. (c-e): The dependence of the clustering coefficient on the node's degree in three organisms: *Aquidex Aeolicus* (archaea) (c), *Escherichia coli* (bacterium) (d), and *S. cerevisiae* (eukaryote) (e). In (f) the $C(k)$ curves averaged over all 43 organisms is shown, while the inset displays all 43 species together. The data points are color coded as above. In (c-f) the dashed lines correspond to $C(k) \sim k^{-1}$, and in (c-e) the diamonds represent $C(k)$ expected for a scale-free network (Fig. 1a) of similar size, indicating the absence of scaling. The wide fluctuations are due to the small size of the network.

*Figure 3.* **Uncovering the underlying modularity of a complex network.**

(a) Topological overlap illustrated on a small hypothetical network. For each pair of nodes, $i$ and $j$, we define the topological overlap $O_T(i,j) = J_n(i,j)/\min(k_i,k_j)$, where $J_n(i,j)$ denotes the number of nodes to which both $i$ and $j$ are linked to (plus one if there is a direct link between $i$ and $j$) and $\min(k_i,k_j)$ is the smaller of the $k_i$ and $k_j$ degrees. On each link we indicate the topological overlap for the connected nodes and in parenthesis next to each node we indicate the node's clustering coefficient. (b) The topological overlap matrix corresponding to the small network shown in (a). The rows and columns of the matrix were reordered by the application of an average linkage clustering method *(22)* to its elements, allowing us to identify and place close to each other those nodes that have high topological overlap. The color code denotes the degree of topological overlap between the nodes (see sidebar). The associated tree clearly reflects the three distinct modules built into the model of Fig. 3a, as well as the fact that the *EFG* and *HIJK* modules are closer to each other in topological sense that the *ABC* module.

*Figure 4.* **Identifying the functional modules in *E. coli* metabolism.**

(a) The topologic overlap matrix corresponding to the *E. coli* metabolism, together with the corresponding hierarchical tree (top) that quantifies the relationship between the different modules. The branches of the tree are color coded to reflect the functional classification of their substrates. The biochemical classes we used to group the metabolites represent carbohydrate metabolism (blue), nucleotide and nucleic acid metabolism (red), protein, peptide and amino acid metabolism (green), lipid metabolism (cyan), aromatic compound metabolism (dark pink), monocarbon compound metabolism (yellow) and coenzyme metabolism (light orange) *(15)*. The color code of the matrix denotes the degree of topological overlap shown in the matrix. On the bottom we show the large-scale functional map of the metabolism, as suggested by the hierarchical tree. (b) 3-D representation of the reduced *E. coli* metabolic network. Each node is color coded by the functional class to which it belongs, and is identical to the color code applied to the branches of the tree shown in (a). Note that the different functional classes are visibly segregated into topologically distinct regions of metabolism. The blue-shaded region denotes the nodes belonging to pyrimidine metabolism, discussed below. (c) Enlarged view of the substrate module of pyrimidine metabolism. The colored boxes denote the three-level, nested modularity suggested by the hierarchical tree. (d) A detailed diagram of the metabolic reactions that surround and incorporate the pyrimidine metabolic module. Red boxes denote the substrates directly appearing in the reduced metabolism and the tree shown in (c). Substrates in green boxes are internal to pyrimidine metabolism, but represent members of non-branching pathways or end pathways branching from a metabolite with multiple connections (*16*). Blue and black boxes show the connections of pyrimidine metabolites to other parts of the metabolic network. Black boxes denote core substrates belonging to other branches of the metabolic tree (a), while blue boxes denote non-branching pathways (if present) leading to those substrates. Note, that with the exception of carbamoyl-phosphate and S-dihydroorotate, all pyrimidine metabolites are connected with a single biochemical reaction. The shaded boxes around the reactions highlight the modules suggested by the hierarchical tree. The shaded blue boxes along the links display the enzymes catalyzing the corresponding reactions, and the arrows show the direction of the reactions according to the WIT metabolic maps (*15*).

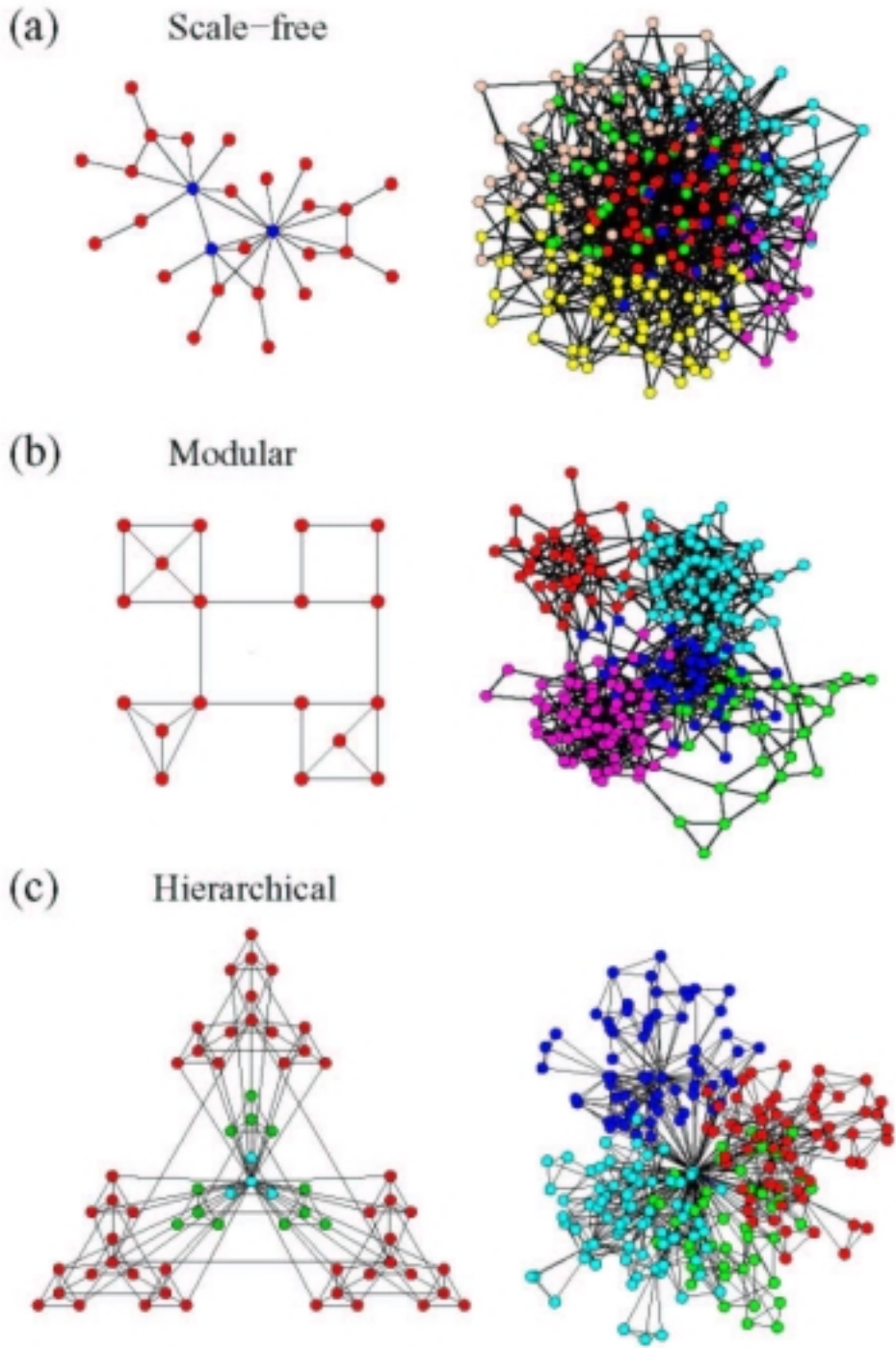

Fig.1, Ravasz et.al.

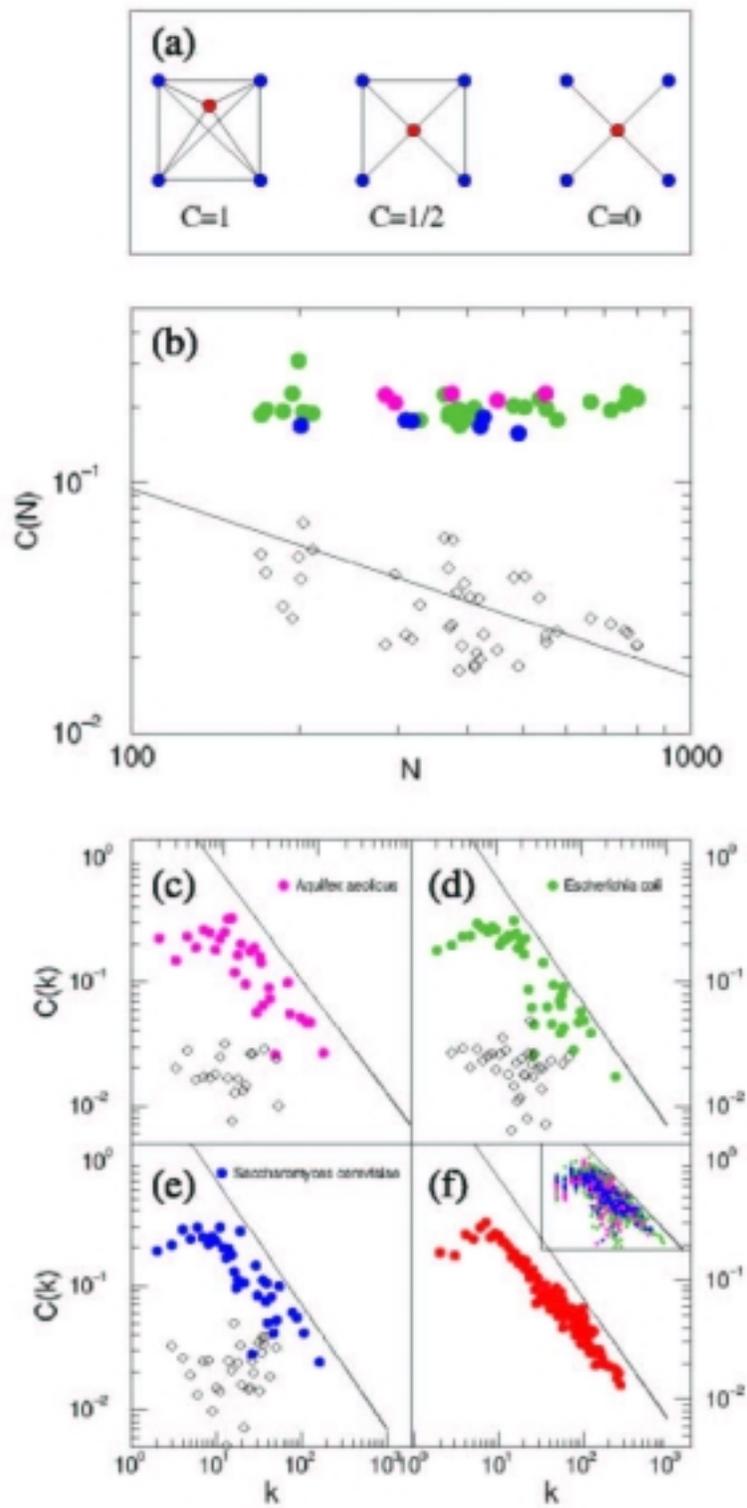

Fig. 2, Ravasz et. al.

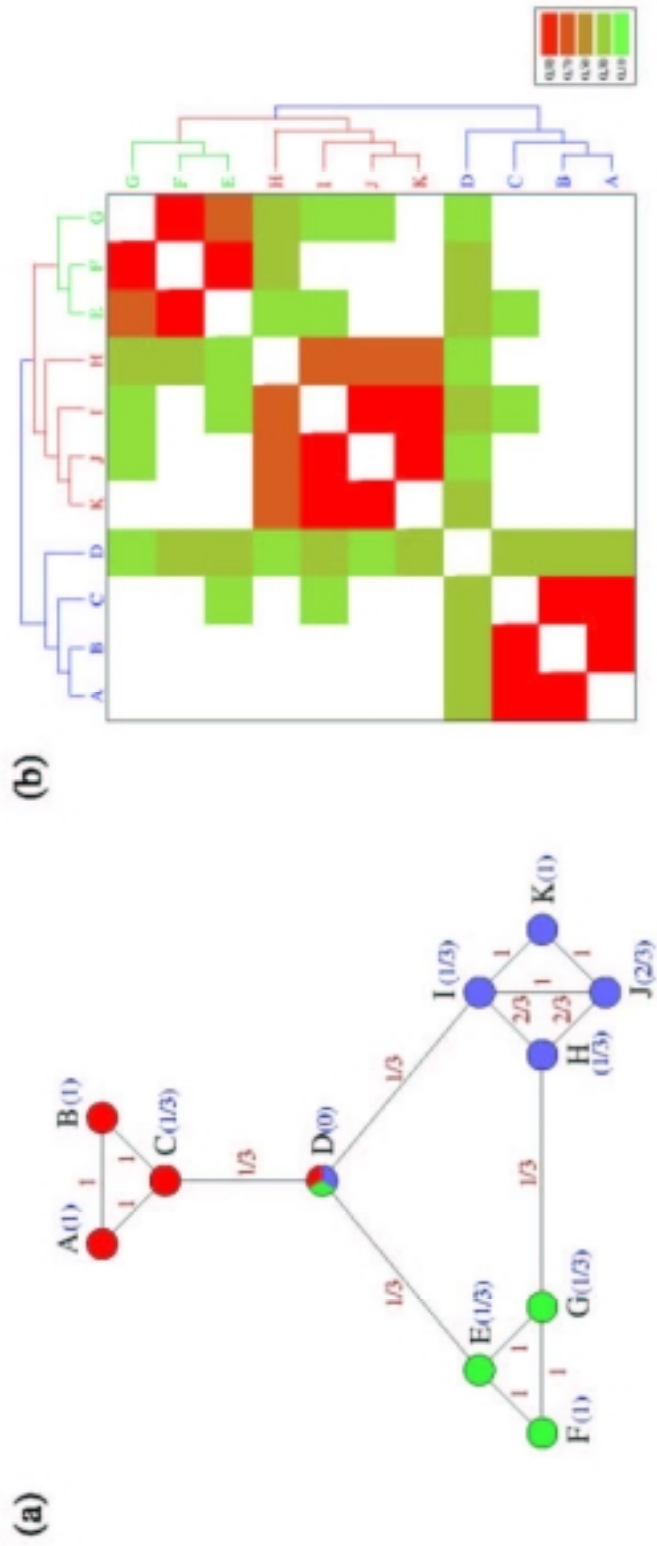

Fig. 3, Ravasz et. al.

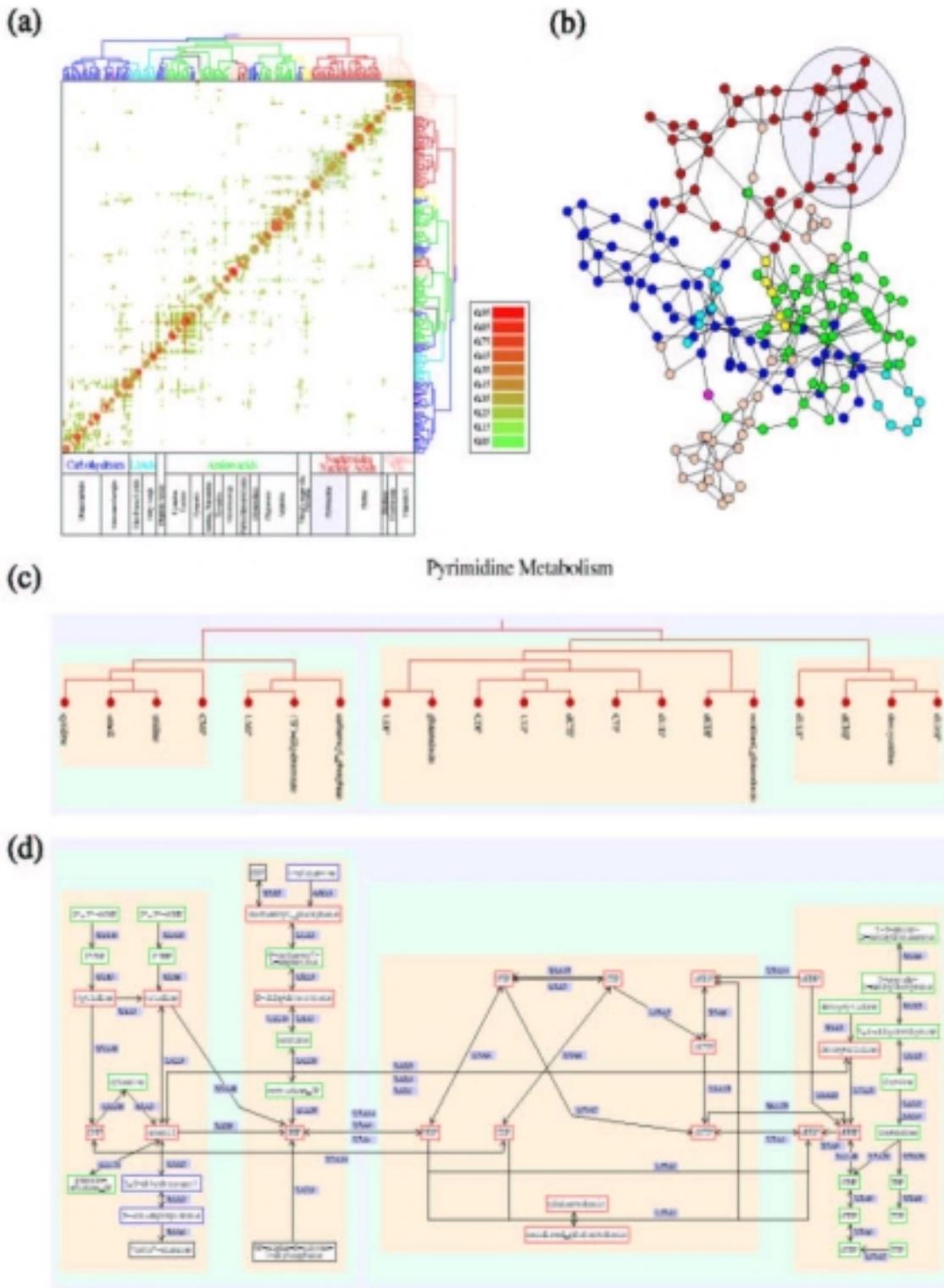

Fig. 4, Ravasz et. al.